\documentclass[10pt,conference]{IEEEtran}

\usepackage{cite}
\usepackage{flushend}
\usepackage{amsmath,amssymb,amsfonts}
\usepackage{algorithmic}
\usepackage{multirow}
\usepackage{graphicx}
\graphicspath{ {./images/} }
\usepackage[braket, qm]{qcircuit}
\usepackage{textcomp}
\usepackage[table]{xcolor}
\usepackage{dirtytalk}
\usepackage{tikz}
\usetikzlibrary{positioning}
\def\BibTeX{{\rm B\kern-.05em{\sc i\kern-.025em b}\kern-.08em
    T\kern-.1667em\lower.7ex\hbox{E}\kern-.125emX}}
    
\begin{document}

\title{HAMMR-L: Noise Reduction in Quantum Outcomes Using a Richardson-Lucy Deconvolution Algorithm for Quantum State Graphs}

\author{
	\IEEEauthorblockN{Jake Scally}
	\IEEEauthorblockA{\textit{Physics Department} \\
	\textit{Florida State University}\\
	Tallahassee, USA \\
	jscally@fsu.edu}
	\and
	\IEEEauthorblockN{Austin Myers}
	\IEEEauthorblockA{\textit{Mathematics Department} \\
	\textit{Florida State University}\\
	Tallahassee, USA \\
	atm22b@fsu.edu}
	\and[\hfill\mbox{}\par\mbox{}\hfill]
	\IEEEauthorblockN{Ryan Carmichael}
	\IEEEauthorblockA{\textit{Computer Science Department} \\
	\textit{Florida State University}\\
	Tallahassee, USA \\
    rcc22e@fsu.edu \\
	} 
	\and
	\IEEEauthorblockN{Phat Tran}
	\IEEEauthorblockA{\textit{Computer Science Department} \\
	\textit{Florida State University}\\
	Tallahassee, USA \\
	ttran@fsu.edu}
	\and
	\IEEEauthorblockN{Xiuwen Liu}
	\IEEEauthorblockA{\textit{Computer Science Department} \\
	\textit{Florida State University}\\
	Tallahassee, USA \\
	xliu@fsu.edu}
}

\maketitle

\begin{abstract}
Current quantum computers present significant noise, especially as circuit depth and qubit count increase. Prior work has demonstrated that erroneous outcomes exhibit some behavior in Hamming space, enabling improvements in the output distributions of NISQ-era computers. We present HAMMR-L: a principled post-processing technique for improving the fidelity of output distributions by applying Richardson-Lucy image deconvolution on a state graph of measurement results connected by Hamming distance. We show that this preliminary implementation of HAMMR-L outperforms existing cutting-edge Hamming-based post-processors such as QBEEP while being circuit and hardware agnostic, which QBEEP is not. HAMMR-L also demonstrates clear potential for future improvements and we discuss how such improvements might be realized while highlighting the strengths, limitations, and generality of the underlying concept.
\end{abstract}


\section{Introduction}
As the field of quantum computing continues to grow, qubit count increases rapidly\cite{ibmroadmap}\cite{googleroadmap}\cite{queraroadmap}. While researchers look towards creating logical qubits with quantum error correction (QEC), realizing useful systems with error-corrected logical qubits is likely to be years down the road\cite{googleroadmap}\cite{queraroadmap}. In order to make our current noisy intermediate-scale quantum (NISQ)-era quantum computers more directly impactful to users today and tomorrow, we turn to quantum error mitigation (QEM). Instead of fully correcting for errors, we seek to reduce the noise in the output distributions of today's quantum computers.

There are many techniques and approaches in QEM that have shown promise\cite{qem}. Within recent years, researchers have observed that output distributions exhibit patterns in \textit{Hamming space}\cite{hammer}\cite{qbeep}\cite{hammingvis}. Hamming space, explained further in Section~\ref{subsec:hamming}, is essentially a space where output strings are represented not by their actual value, but instead their distance (or number of bit-flip errors) from a correct answer. Due to the observed structures in Hamming space, researchers have raised the question of whether the behavior can be leveraged to mitigate errors. This has shown promise in \cite{hammer} and \cite{qbeep}, where the authors were able to demonstrate two different Hamming-based methods that successfully improve output distributions.

Building on the success shown in this field, we present a principled approach to QEM using Hamming structure and Richardson-Lucy deconvolution. We demonstrate that by applying Richardson-Lucy deconvolution with a Hamming-based blurring function, HAMMR-L provides greater improvement in output distributions than QBEEP\cite{qbeep}, the current leader in Hamming-based quantum error mitigation.

\section{Background}
\subsection{Bernstein-Vazirani Algorithm for QEM Evaluation}

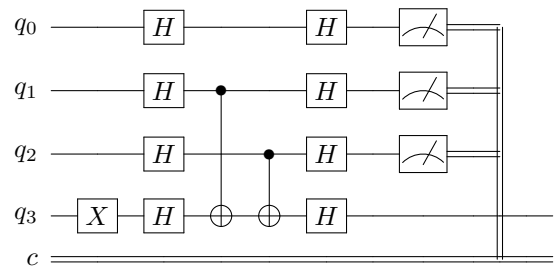
\begin{figure}[h]
    	\centering
	\[
	\Qcircuit @C=1em @R=1em {
    		\lstick{q_0} & \qw      & \gate{H} & \qw      & \qw      & \gate{H} & \qw      & \meter & \cw & \cw \cwx[4] & \\
    		\lstick{q_1} & \qw      & \gate{H} & \ctrl{2} & \qw      & \gate{H} & \qw      & \meter & \cw & \cw & \\
    		\lstick{q_2} & \qw      & \gate{H} & \qw      & \ctrl{1} & \gate{H} & \qw      & \meter & \cw & \cw & \\
    		\lstick{q_3} & \gate{X} & \gate{H} & \targ    & \targ    & \gate{H} & \qw      & \qw    & \qw & \qw & \qw & \qw \\
    		\lstick{c} & \cw & \cw & \cw & \cw & \cw & \cw & \cw & \cw & \cw & \cw & \cw
	}
	\]
	\caption{BV example circuit for the secret string ``110.'' The classical register $c$ should hold exactly ``110'' with certainty on a fault-tolerant quantum computer. However, this is not the case in practice.}
	\label{fig:bv-example}
\end{figure}

The Bernstein-Vazirani (BV) algorithm is frequently used for illustrating results in QEM, including in HAMMER\cite{hammer} and QBEEP\cite{qbeep}. While the BV problem is not hugely useful in many ways, it is a great illustrative example due to the circuit's many entangling operations, two-qubit gates, and single correct answer. The BV algorithm shows that a quantum computer can determine an \textit{oracle function}'s secret string in a single query. There is extensive literature\cite{bv} explaining how this works, but for the purposes of this paper we will highlight a few important features. Figure~\ref{fig:bv-example} shows how the BV algorithm is implemented as a quantum circuit. Although this example is fairly trivial, it still illustrates that the circuit returns ``110'' with certainty and has a number of CNOT gates corresponding to the number of ones in the secret string. This allows for some control over the error rates of the circuit (for example, ``11111'' will have a higher error rate than ``10101'' due to there being more CNOTs), making it easier to evaluate HAMMR-L's effectiveness for a variety of error rates.

\subsection{Hamming Space} \label{subsec:hamming}
\textit{Hamming distance} refers to the number of different symbols between two strings. For example, the strings ``1111'' and ``1011'' have a hamming distance of 1. Hamming distance is essentially a way of quantifying how far off an incorrect result is from a correct one.

Given some $N\in \mathbb{N},$ the $N$-\textit{Hamming space} $H_N$ is the metric space formed on the set of all binary strings $a_1a_2\cdots a_N,$ $a_i\in\{0,1\}$ of length $N.$ The metric on this set is the \textit{Hamming distance} of two strings $x,y\in H_N$, denoted $h(x,y),$ and is defined as the number of digits for which the strings $x$ and $y$ differ. Note that $H_N$ always contains $2^N$ strings. Given an $N$-qubit quantum circuit $C$, the output of $C$ will be an element of $H_N.$

Prior works use Hamming space to estimate the correct output by utilizing local information around each observed point in $H_N$, a feature that we also exploit in HAMMR-L\cite{hammer}\cite{qbeep}\cite{hammingvis}.

\subsection{Previous Hamming-Based QEM Implementations}
In \cite{hammer}, the HAMMER algorithm utilizes Hamming space to post-process the output distribution of a quantum circuit. The goal is to assign a \say{neighborhood score} $S(x)$ to each $x\in H_N$. HAMMER assigns a weight to each neighbor a fixed distance away, sums over them, and then updates the probability of each $x$ to be $Pr(x)\cdot S(x).$ This is done by creating a state graph of the outcomes connected to their respective neighbors in Hamming space. 

Given $x\in H_N$ and $0\leq i \leq N/2$, the \textit{cumulative Hamming strength} $CHS_i(x)$ is the total probability of all strings $y$ at a fixed distance $i$ away from $x,$ $$CHS_i(x)=\sum_{h(x,y)=i}Pr(y)$$ Using the CHS, define weights 
\[
W_i(x)= \begin{cases}
\frac{1}{CHS_i(x)} & \text{if \hspace*{2mm} } CHS_i(x)\neq 0\\
0 & \text{if \hspace*{2mm} }  CHS_i(x)=0.
\end{cases}
\]

Then, define a filter function $\pi(x,y)$ which only considers neighbors with probability \textit{lower} than $x:$ 
\[
\pi(x,y)= \begin{cases}
1 & \text{if \hspace*{2mm} } Pr(y) < Pr(x)\\
0 & \text{if \hspace*{2mm} } Pr(y) \geq Pr(x).
\end{cases}
\]

In principle, this filter function $\pi$ avoids boosting incorrect outcomes with low probabilities that happen to be in an \say{influential} neighborhood in which neighbors may have high probabilities.

The final neighborhood score $S(x)$ is given by $$S(x)=\sum_{i=1}^{\lfloor N/2\rfloor}W_i(x)\cdot\sum_{y\text{ s.t. } d(x,y)=i}Pr(y)\cdot\pi(x,y),$$ where the second sum is again taken over $y\in H_N$ such that $d(x,y)=i.$ Each string $x\in H_N$ is then updated with a \textit{likelihood} function given by $$L(x)=S(x)\cdot Pr(x).$$

HAMMER first presented this idea of building a state graph connected in Hamming space to help mitigate errors. QBEEP takes this a step further and constructs the same underlying state graph, but instead uses a Poisson distribution for edge weighting. The shape of the Poisson distribution in QBEEP is tailored to the circuit itself and the quantum hardware error rates in order to more accurately describe the error distribution. QBEEP is shown to outperform HAMMER in \cite{qbeep}, so we later quantify HAMMR-L's performance directly against QBEEP.

\section{HAMMR-L Implementation}

\subsection{Graph Construction}
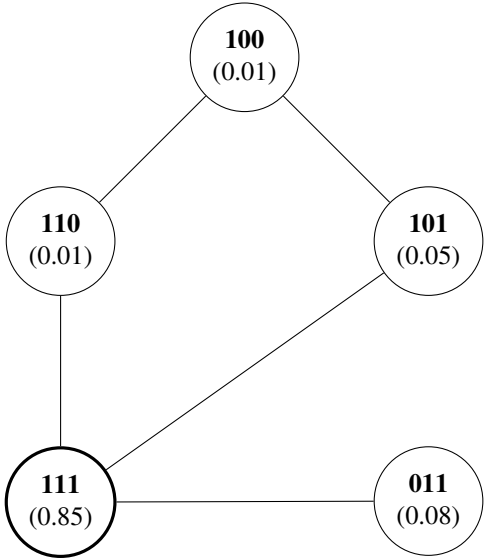
\begin{figure}
	\begin{center}
		\caption{Example state graph based on experimental results for the 3-qubit Bernstein-Vazirani circuit for a secret string ``111.'' Nodes have their observed string and observed probabilities based on the number of times that the string was observed divided by the total shots. Nodes are connected to their neighbors in Hamming space, meaning that nodes one edge away are a single bit flip from the source node, two edges away are two bit flips from the source node, and so on.}
		\vspace{5mm}
		\label{fig:stategraphexample}
		\begin{tikzpicture}[node distance=2cm, every node/.style={draw, circle, align=center}]
    			\node (A) {\textbf{100}\\ \textnormal{(0.01)}};
    			\node (B) [below left=of A] {\textbf{110}\\ \textnormal{(0.01)}};
    			\node (C) [below right=of A] {\textbf{101}\\ \textnormal{(0.05)}};
    			\node[very thick] (D) [below =of B] {\textbf{111}\\ \textnormal{(0.85)}};
    			\node (E) [below =of C] {\textbf{011}\\ \textnormal{(0.08)}};

    			\draw (A) -- (B);
    			\draw (A) -- (C);
    			\draw (B) -- (D);
    			\draw (C) -- (D);
    			\draw (D) -- (E);

		\end{tikzpicture}
	\end{center}
\end{figure}

HAMMR-L employs the same initial step as both HAMMER and QBEEP. A state graph of the measured results is constructed with each node being an observed string and the percentage of the total counts it had. Nodes are connected to each other if their strings have a distance of 1 in Hamming space (see Figure~\ref{fig:stategraphexample}).

\subsection{Richardson-Lucy Deconvolution} \label{subsec:richardsonlucy}
Richardson-Lucy (R-L) deconvolution was initially an image deblurring algorithm used for improving the clarity of detectors\cite{richardson}\cite{lucy}. It has since been used in all kinds of image and photography software, including the popular open-source application RawTherapee. R-L deconvolution operates under the assumption that there is an underlying image that has been blurred by some point spread function (PSF). This PSF dictates how much a single ``lit up'' pixel would spread out across neighboring pixels in an image. If this PSF is known, the deconvolution converges to the maximum likelihood solution for the original image\cite{maxlikelihoodR-L}.

Though this algorithm is more frequently used for two-dimensional data like images, the process is the same regardless of dimensionality. As such, consider Hamming space. For a circuit of $n$ qubits, the Hamming space is $n$ dimensional. We can take the state graph connected by distance in Hamming space (Figure~\ref{fig:stategraphexample}) and use it as our ``image.'' By treating the probabilities of each node as the ``brightness'' of a particular pixel, we can attempt to deconvolve the original, correct distribution.

Further, consider the meaning of a point spread function. In our case, the PSF indicates how much a single correct string will ``spread out'' into other incorrect strings. The Hamming distance structures observed in \cite{hammer} and \cite{qbeep} show that our PSF should be a function of Hamming distance.

The challenge of mitigating errors in a distribution of output strings is formally quite similar to the original application of deblurring pixels. To see this, consider a function $d(i)$ denoting the observed probability of measuring string $i$, and another function $u(i)$ denoting the \textit{true} probability of observing string $i$. Then the noisy distribution can be viewed as a discrete convolution (in this case, over the finite field $\mathbb{F}_{2^N},$ which can be identified with $H_N$) between the true distribution and a given point spread function $P$ which depends on the Hamming distance between two strings: $$d(i)=(u*P)(i)= \sum_ju(j)P(h(i,j))$$ Note that the standard distance on $\mathbb{R}$ is denoted $|x-y|,$ so we use $h(i,j)$ in the argument of the PSF in the convolution to represent the distance in Hamming space ($H_N$). Thus, in order to recover the original distribution, we must to deconvolve the blurred probability distribution $d(i).$

For the sake of notational clarity, we will now suppress the function notation and express $d_i:=d(i)$ as the probability of the $i^{th}$ string. Further, we also express $u_i:=u(i)$ as the true (error-free) probability of the $i^{th}$ string, and take $$p_{ij}:=\frac{1}{h(i,j)+1}$$ to be the PSF evaluated at $h(i,j),$ the distance between the $i^{th}$ and $j^{th}$ strings. Thus far, we have experimented with various dependencies on Hamming distance with varying success, which will be discussed in Section~\ref{sec:discussion}. We then, much like HAMMER and QBEEP, organize Hamming space into a weighted graph with a node for each experimentally observed string $i\in H_N,$ and the corresponding probability $d_i$ to each node. We then attach an edge between each pair of nodes $(i,d_i)$ and $(j, d_j)$ such that $h(i,j)=1.$ To estimate each true probability $u_i,$ the standard Richardson-Lucy deconvolution procedure is performed. Let $\hat{u}_{j}^{(n)}$ denote the $n^{th}$ estimate of $u_j,$ the true probability distribution, and set $$\hat{u}_j^{(0)}:=d_j,$$ $$\hat{u}_j^{(n+1)}:=\hat{u}_j^{(n)}\sum_{\substack{i}}\frac{d_i}{c_i}p_{ij}$$ 
with $$c_i:=\sum_{\substack{j}}p_{ij}u_j^{(n)}$$ and both sums being taken over all other nodes in the graph (this can, however, be cut off for very high-qubit circuits). Once the algorithm has converged after $n$ iterations, it is possible that we no longer have $\sum_{\substack{i}} u_i^{(n)}=1,$ so we normalize each $u_i^{(n)}$ and get the final estimated probabilities $$\hat{u}_i:=\frac{u_i^{(n)}}{\sum_{\substack{i}}u_i^{(n)}}$$ This process is repeated until convergence or until a maximum number of iterations is reached.

\begin{figure}
	\includegraphics[width=\linewidth]{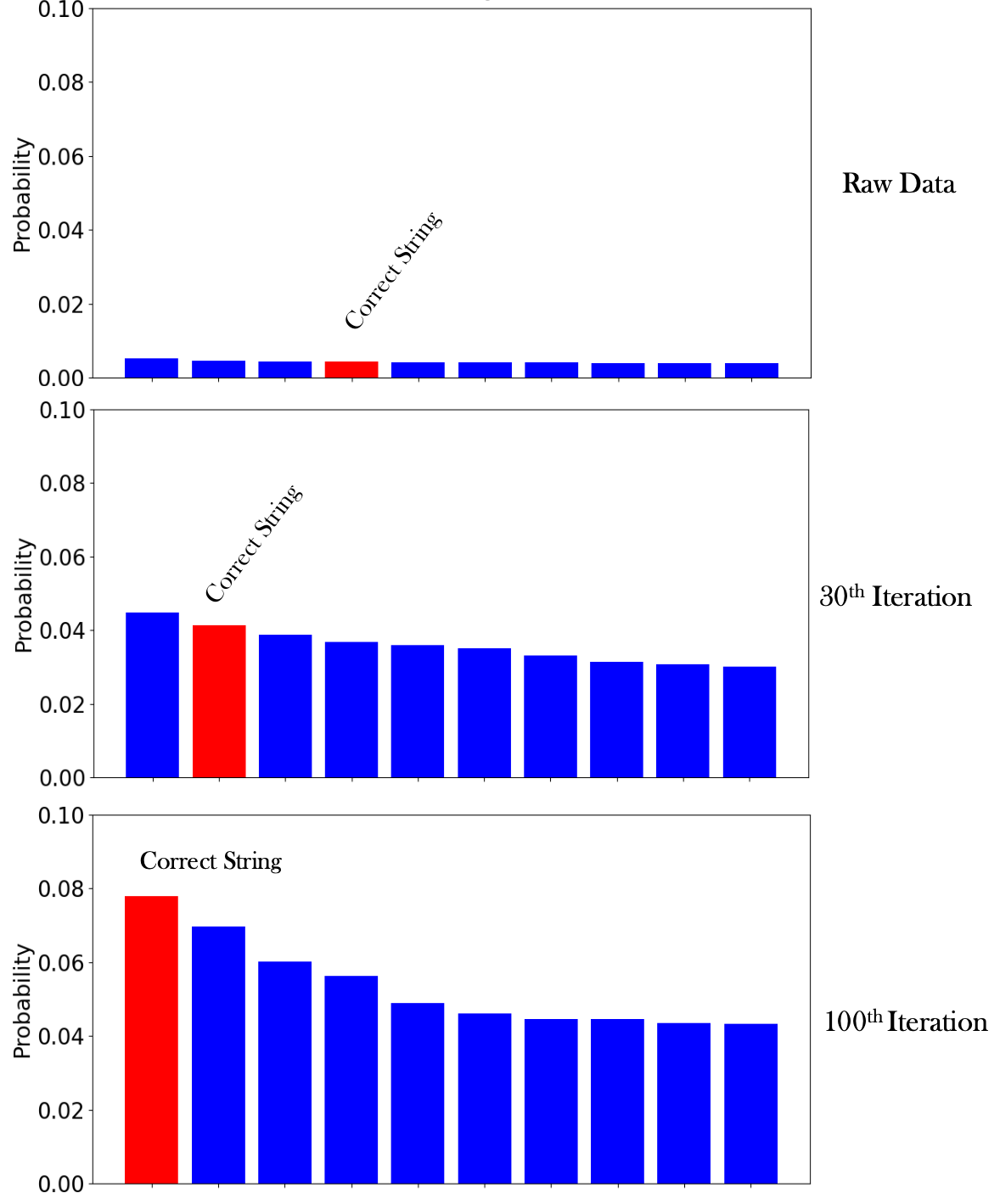}
	\caption{Example showing HAMMR-L on the top ten most likely strings for the nine-qubit Bernstein-Vazirani circuit for the secret string ``111111111.'' As this is a circuit with high entanglement and many CNOTs, the output is hardly usable. HAMMR-L is able to increase the rank from 4th to 1st in 100 convolutions and increase the probability from less than 1\% to nearly 8\%.}
	\label{fig:9BVex}
\end{figure}

\begin{figure}
	\includegraphics[width=\linewidth]{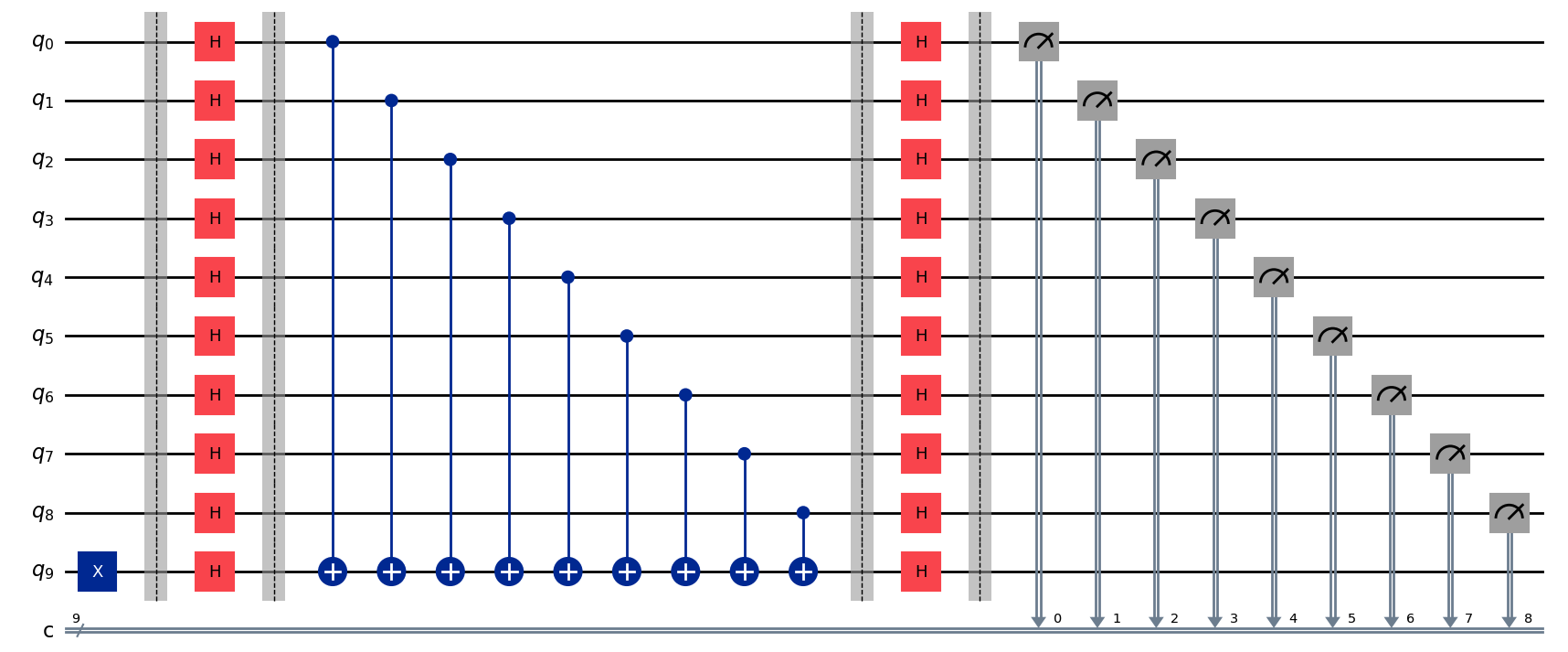}
	\caption{9-qubit Bernstein-Vazirani circuit for the secret string ``111111111.'' This circuit exhibits extremely high error rates on IBM's quantum computers due to the depth, qubit count, number of two-qubit gates, and high entanglement.}
	\label{fig:9BVcircuit}
\end{figure}

\section{Experimental Techniques}

Due to the high degree of variability on the IBM quantum computers, it is more illustrative of the actual performance to generate large datasets of circuits with similar error rates as opposed to a single specific circuit. To do this, we use the Bernstein-Vazirani algorithm, picking a qubit count and ``ones count'' in the secret string and then generating and running all circuits corresponding to every permutation of the secret string. The ones count is simply how many 1s are in the secret string. For example, our nine-qubit six-ones dataset contains BV circuits for ``000111111,'' ``001111110,'' ``001111101,'' etc. Due to the way these circuits are constructed, the number of ones corresponds to the number of CNOTs and hence the error rate. This gives us access to a variety of error rates just by tuning the ones count, while running all of the permutations attempts to control for bad qubits and random fluctuations in circuit execution.

We have generated nine-qubit datasets for four, five, six, and seven ones and applied HAMMR-L and QBEEP on the results of each. We also collected a ten-qubit eight-ones dataset, using ten qubits as opposed to nine qubits due to there being few permutations of eight-ones strings with just nine qubits. The four-ones case has high signal and acts as a test to ensure that our algorithm does not hurt results that are already correct. The eight-ones case has very little obvious signal, but HAMMR-L and QBEEP are still able to recover the correct result in some cases. Further, we collected each dataset twice in an effort to observe the consistency between runs. In the future, we hope to collect more, but we were only able to run each dataset twice due to quantum hardware compute time restrictions.

With these datasets, we convert the counts to a probability by dividing the counts by the shots, or number of executions of each circuit. We then apply HAMMR-L and QBEEP to the datasets. Both HAMMR-L and QBEEP tend to increase the probability of the correct string, but this does not illustrate how each algorithm performs in improving the probability of the correct string \textit{while also} decreasing the probability of incorrect strings. As such, we primarily look at the change in \textit{rank}. The highest probability is ranked 1st, second highest probability is ranked 2nd, and so on. We measure three main categories to evaluate the performance of HAMMR-L and QBEEP:
\begin{center}
	\begin{tabular}{|c|c|c|}
		\hline
		Category & Example & Ex. Rank Change\\ \hline
		Rank improved & 4th to 2nd & 2 \\
		Rank unchanged & 4th to 4th & 0 \\
		Rank worsened & 4th to 7th & -3 \\ \hline
	\end{tabular}
\end{center}
These metrics provide a more meaningful measure of how useful each post-processing technique is and how they affect the probability of the correct string compared to incorrect strings.

\section{Results}
\label{sec:results}

\begin{table*}[h]
	\centering
	\renewcommand{\arraystretch}{1.5}
	\caption{Performance of HAMMR-L and QBEEP on nine-qubit Bernstein-Vazirani circuits with a variety of secret strings\\ (The eight-ones set was run with 10 qubits to allow for more permutations)}
	\label{tab:alldata}
	\begin{tabular}{|c|c|cc|cc|}
		\hline
		& & \multicolumn{2}{c|}{Run 1} & \multicolumn{2}{c|}{Run 2} \\
		Secret String & Post-processed Rank Change & HAMMR-L & QBEEP & HAMMR-L & QBEEP \\
		\hline
		\multirow{4}{*}{Four ones} & Rank Improved & \cellcolor{green!30}4.8\% & \cellcolor{red!30}4.0\% & \cellcolor{green!30}0.8\% & \cellcolor{red!30}0.0\% \\
		                                          & Rank Unchanged & 92.8\% & 93.6\% & 98.4\% & 100\% \\
		                                          & Rank Worsened & \cellcolor{yellow!30}2.4\% & \cellcolor{yellow!30}2.4\% & \cellcolor{red!30}0.8\% & \cellcolor{green!30}0.0\% \\ \cline{2-6}
		                                          & Mean Rank Change & 0.06 & 0.02 & 0 & 0 \\ \hline
		\multirow{4}{*}{Five ones} & Rank Improved & \cellcolor{red!30}4.8\% &\cellcolor{green!30} 9.5\% & \cellcolor{red!30}2.4\% & \cellcolor{green!30}4.0\% \\
							& Rank Unchanged & 69.8\% & 77.8\% & 72.2\% & 77.0\% \\
							& Rank Worsened & \cellcolor{red!30}28.6\% & \cellcolor{green!30}12.7\% & \cellcolor{red!30}25.4\% & \cellcolor{green!30}19.0\% \\ \cline{2-6}
							& Mean Rank Change & -0.7 & -0.4 & -1.0 & -0.6 \\ \hline
		\multirow{4}{*}{Six ones} & Rank Improved & \cellcolor{green!30}38.1\% & \cellcolor{red!30}22.6\% & \cellcolor{green!30}34.5\% & \cellcolor{red!30}27.4\% \\
						      & Rank Unchanged & 41.7\% & 48.8\% & 34.5\% & 36.9\% \\
						      & Rank Worsened & \cellcolor{green!30}20.2\% & \cellcolor{red!30}28.6\% & \cellcolor{green!30}31.0\% & \cellcolor{red!30}35.7\% \\ \cline{2-6}
						      & Mean Rank Change & 4.6 & -0.6 & 2.5 & -2.9 \\ \hline
		\multirow{4}{*}{Seven ones} & Rank Improved & \cellcolor{red!30}38.9\% & \cellcolor{green!30}41.7\% & \cellcolor{green!30}58.3\% & \cellcolor{red!30}36.1\% \\
							   & Rank Unchanged & 22.2\% & 22.2\% & 8.4\% & 27.8\% \\
							   & Rank Worsened & \cellcolor{red!30}38.9\% & \cellcolor{green!30}36.1\% & \cellcolor{green!30}33.3\% & \cellcolor{red!30}36.1\% \\ \cline{2-6}
							   & Mean Rank Change & -1.75 & -0.5 & 3.9 & -1.1 \\ \hline
		\multirow{4}{*}{Eight ones (10 qubit)} & Rank Improved & \cellcolor{green!30}66.7\% & \cellcolor{red!30}51.1\% &\cellcolor{green!30} 55.6\% & \cellcolor{red!30}44.4\% \\
									& Rank Unchanged & 8.9\% & 31.1\% & 6.7\% & 37.8\% \\
									& Rank Worsened & \cellcolor{red!30}24.4\% & \cellcolor{green!30}17.8\% & \cellcolor{red!30}37.7\% & \cellcolor{green!30}17.8\% \\ \cline{2-6}
									& Mean Rank Change & 10.7 & 2.2 & 12.5 & 0.6 \\ \hline
	\end{tabular}
\end{table*}
When post-processing our datasets with HAMMR-L and QBEEP, HAMMR-L demonstrates a higher percentage of Rank Improved circuits than QBEEP in seven of the ten datasets. Further, HAMMR-L demonstrates a \textit{significant} improvement in mean rank increase. Averaged across all ten datasets, HAMMR-L has an average rank increase of 3.081, while QBEEP has an average rank change of -0.328, indicating that QBEEP on average worsened the rank of the correct string. The post-processed rank data for these datasets are shown in Table~\ref{tab:alldata}.

Figure~\ref{fig:9BVex} is an excellent demonstration of HAMMR-L's signal recovery ability. For a quantum circuit run for 10,240 shots on IBM Brisbane, the correct string ``111111111'' was initially the fourth highest ranked string, with three other erroneous strings having more counts. Further, the probabilities for all strings were quite low due to the challenging nature of the circuit (see Figure~\ref{fig:9BVcircuit}). In spite of this, HAMMR-L demonstrates its ability to find the correct solution among seemingly structureless data. 

Of further note is HAMMR-L's performance on both runs of the six-ones datasets. In both instances, QBEEP had more Rank Worsened result distributions than Rank Improved, a negative overall effect on the rank. In contrast, HAMMR-L had a positive overall effect which was larger than QBEEP's by a substantial margin. However, looking at the five-ones sets, we see that QBEEP outperforms HAMMR-L, although neither had a positive effect on the rank of the secret string. That isn't to say both algorithms did not have a positive effect on the overall distribution, since they still increased the probability of the correct string in many distributions, even though neither were able to selectively improve the secret string's rank compared to other high-probability erroneous strings. The four-ones datasets have minimal information of note other than the fact that neither HAMMR-L nor QBEEP negatively affected the rank of the secret strings by a substantial amount. Note that many of these circuits already had the correct string ranked 1st, so rank improvements are inherently more infrequent.

In the seven-ones dataset, HAMMR-L demonstrated high run-to-run variance, underperforming QBEEP in Ranks Improved on Run 1 and outperforming QBEEP by over 20\% in Run 2. The seven-ones datasets demonstrated by far the most run-to-run variance for HAMMR-L. Currently, we are not able to deduce the reason for the poor performance on Run 1, but we suspect that it was due to a fluctuation or re-calibration in IBM's quantum hardware leading to error rates that did not align well with our catch-all PSF. It is likely that the future improvements discussed in Section~\ref{sec:discussion} could help with hardware variations.

HAMMR-L excelled in the eight-ones datasets, improving the rank in over half the results across both runs. Further, the mean rank change was very high for this dataset, over eight points higher than QBEEP in both runs. We note, however, that QBEEP had fewer Rank Worsened distributions than HAMMR-L. This is the only dataset where we observed a significant disconnect between the ``winner'' in the Rank Improved and Rank Worsened categories.

\section{Discussion and Future Work}
\label{sec:discussion}
With these results, HAMMR-L seems to outperform prior work in this field. However, we would like to continue evaluating HAMMR-L's performance on different circuits and quantum computers. Due to quantum hardware compute time limitations, we were only able to collect the data presented in this paper. While this data shows strong performance for HAMMR-L, the run-to-run variance is high. More work is needed in the evaluation of HAMMR-L to further understand its performance and look for behavior that could lead to future improvements.

That said, the HAMMR-L framework shows clear promise in terms of error mitigation performed using Richardson-Lucy beyond the preliminary work done in this paper. In particular, the PSF representation of $\frac{1}{h(i,j)}$ was found experimentally. We tried using various Hamming distance dependencies (such as Poisson and binomial distributions, as suggested by QBEEP's results\cite{qbeep}) and found that $\frac{1}{h(i,j)}$ (as used in HAMMER\cite{hammer}) generally worked the best. However, Richardson-Lucy deconvolution relies on the PSF being as accurate as possible, suggesting that improving on this PSF could greatly improve the results of this technique.

We see two main ways of approaching this. The first is an approach tailored to experimental circuit and hardware error rates, as done in QBEEP\cite{qbeep}. Noise models for the IBM quantum computers are publicly available and could be used in conjunction with the layout of the circuit being run to create an accurate estimate of how the errors will distribute themselves in Hamming space. Although this would likely produce a very accurate PSF, it would be hardware and circuit layout dependent which would add significant overhead and rely on accurate reporting from quantum computer providers for success.

The second approach, which we believe is the ``next step'' for HAMMR-L, is blind deconvolution. Blind deconvolution is an approach to image recovery in which the PSF is not known and is instead estimated based on the structure of the blurred image. This field has been the subject of extensive research spanning multiple decades, but many approaches to blind deconvolution are outlined well in \cite{blinddeconvolution}. These approaches are likely to scale well to the problem of quantum error mitigation and we hope to explore them in the future.

Finally, there is another option in which Richardson-Lucy deconvolution is combined with deep learning\cite{richlucynetwork}\cite{lucyd}\cite{deepurl}. These approaches, although differing slightly, all demonstrate substantial improvements over classical Richardson-Lucy deconvolution, often being able to learn the PSF itself. As such, we expect that this family of methods could also demonstrate improvements in HAMMR-L in both speed and accuracy.
\section{Conclusion}
In this work, we present HAMMR-L: a novel post-processing technique for error mitigation in NISQ-era quantum computers using Richardson-Lucy deconvolution on quantum state graphs connected by Hamming distance. Through experimental results of the Bernstein-Vazirani algorithm on IBM quantum hardware, HAMMR-L demonstrates measurable improvements over prior Hamming-based techniques such as QBEEP, especially in high-error regimes, while remaining circuit and hardware agnostic.

While the demonstrated rank improvements of HAMMR-L already show practical promise, our method provides a principled and generalizable framework for ``deblurring'' quantum output distributions, backed by well-established image processing tools. We believe that there are multiple paths forward for Richardson-Lucy-based output deconvolution, including hardware-aware PSF modeling and blind deconvolution. 

In conclusion, HAMMR-L represents a step forward in quantum error mitigation techniques and a step towards making the NISQ-era quantum computers more useful through the Richardson-Lucy framework.

\bibliography{references}

\end{document}